\documentclass[final,times,twocolumn]{elsarticle}
\usepackage{graphicx,amssymb,amsmath,hyperref}
\usepackage{multirow,dcolumn,bm,latexsym,soul}

\journal{Physics of the Dark Universe}
\begin{document}
\newcommand{\be}{\begin{equation}}
\newcommand{\ee}{\end{equation}}
\newcommand{\bq}{\begin{eqnarray}}
\newcommand{\eq}{\end{eqnarray}}
\begin{frontmatter}

\title{Varying fundamental constants cosmography}
\author[inst1,inst2]{C. J. A. P. Martins\corref{cor1}}\ead{Carlos.Martins@astro.up.pt}
\address[inst1]{Centro de Astrof\'{\i}sica da Universidade do Porto, Rua das Estrelas, 4150-762 Porto, Portugal}
\address[inst2]{Instituto de Astrof\'{\i}sica e Ci\^encias do Espa\c co, CAUP, Rua das Estrelas, 4150-762 Porto, Portugal}
\cortext[cor1]{Corresponding author}

\begin{abstract}
Cosmography is a model-independent phenomenological approach to observational cosmology, relying on Taylor series expansions of physical quantities as a function of the cosmological redshift or other analogous variables. A recent work developed the formalism for a cosmographic analysis of astrophysical and local measurements of the fine-structure constant, $\alpha$, and provided first constraints on the corresponding parameters. Here we update the earlier work, both by including more recent measurements of $\alpha$, and by extending it to other fundamental dimensionless couplings. We find no statistically significant evidence for such variations, and place stringent constraints on the first two terms of all these cosmographic series: the linear coefficient in the $\alpha$ series is constrained to parts per billion level, thanks to recently improved atomic clock constraints, while the other coefficients are constrained to parts per million level. Additionally, we use the same data to place cosmographic constraints on models from a broad class of Grand Unified Theories in which varying fundamental constants occur, and highlight how future data can provide a discriminating cosmographic test between freezing and thawing dark energy models.
\end{abstract}
\begin{keyword}
Observational cosmology \sep Cosmography \sep Varying fundamental constants \sep Atomic clocks
\end{keyword}
\end{frontmatter}

\section{Introduction}
\label{introd}

One of the core goals of observational cosmology is to map the expansion history of the universe. The traditional approach to this is to choose a fiducial background cosmology, resulting in a model-dependent mapping, whose parameters can be constrained by thr available data. However, there are two alternatives which, at least in principle, are model-independent. The first alternative approach is the redshift drift of objects following the cosmological expansion, also known as the Sandage test \cite{Sandage}. This has not yet been detected and has only been attempted with archival data (a first dedicated experiment is ongoing, using the ESPRESSO spectrograph \cite{Trost}), but is well within reach of the next generation of astrophysical facilities, specifically the ELT and the SKAO \cite{Liske,Klockner,Rocha}. The second alternative approach is known as cosmography \cite{Visser}, and relies on expanding physical quantities of interest as a Taylor series, either in the cosmological redshift or in related variables, around the present time.

Usually the cosmographic approach is applied to the scale factor or to quantities directly derived from it, such as the Hubble parameter. A recent work \cite{Cosmography} applied it to the mapping of possible time---and therefore redshift---variations of the fine-structure constant, $\alpha=e^2/(\hbar c)$, which are expected in many extensions of the standard cosmological model---see \cite{ROPP,Uzan} for recent reviews. Here we build upon the earlier work, both by updating the constraints on the $\alpha$ cosmographic series (benefiting from newly published astrophysical and local measurements) but also by extending it to include cosmographic series for two other astrophysically relevant dimensionless couplings: the proton-to-electron mass ratio $\mu=m_p/m_e$ and the proton gyromagnetic ratio $g_p$. One of the vantages of this extension is that it enables the use of a further 66 local and astrophysical measurements in the analysis, as well as to broaden it to include up to six free parameters. Generically, putative redshift dependencies of a dimensionless fundamental constant $Q(z)$ are expressed relative to their present-day laboratory value $Q_0$, specifically,
\be\label{defa}
\frac{\Delta Q}{Q}(z)\equiv\frac{Q(z)-Q_0}{Q_0}\,;
\ee
in the present work, $Q$ can be $\alpha$, $\mu$, $g_p$ or various multiplicative combinations of the three.

One further reason to expect that a cosmographic approach to varying fundamental constants is useful (beyond the obvious one of model independence) is that any such variations are known to be small, so a Taylor series is seemingly appropriate. Moreover, focusing for a moment on $\alpha$, while its cosmological dependence is of course model-dependent, one knows that in the simplest (best motivated and scalar-field based) models, $\alpha$ is constant during the radiation-dominated era, drifts slowly (typically logarithmically) in the matter-dominated era, and then stabilizes again once the universe starts accelerating, once more motivating a cosmographic approach. Detailed examples of classes of models with this behavior can be found in \cite{Barrow,Damour}. As for the other constants, in many models where $\alpha$ varies they are also expected to vary proportionally to it, with the proportionality factors being different for each constant as well as model-dependent. In the latter part of this work we explicitly consider a class of models which parametrically exhibits this behavior.

One of the traditional premises of the cosmographic approach is that one tends to focus on low redshift data (specifically at redshifts $z<1$), justifying a Taylor series approach and its truncation at a suitably chosen low-order term. This is important since the series can diverge if one includes data at $z>1$. This would certainly be problematic in the case of astrophysical tests of the stability of varying fundamental constants, for which direct measurements already exist up to redshifts $z\sim7$ \cite{Wilczynska}. However, it has been shown \cite{Cattoen} that this divergence can be avoided by expanding in the rescaled redshift, defined as
\be\label{defy}
y\equiv\frac{z}{1+z}\,,
\ee
instead of in the redshift itself. Naturally, the need to carefully choose where the series is truncated  still remains.

Additionally, this approach can be limited by degeneracies between the series coefficients, and a dependence on chosen priors for these coefficients. A general discussion of these points can be found in \cite{Dunsby}, while one focused on the case of the redshift drift is in \cite{Rocha}. Still, as first shown in \cite{Cosmography} and confirmed in what follows, varying constants cosmography is somewhat less vulnerable to these degeneracies than cosmography in the other contexts, and it provides important model-independent constraints which can be converted, for specific models, into relevant constraints on local and cosmological parameters in the models.

An interesting feature of varying fundamental constants cosmography is that it explicitly shows the synegies between the local and astrophysical tests of their stability. Specifically, the linear term in the series is predominantly constrained by laboratory tests using atomic clocks, while the higher-order terms are only constrained by the astrophysical measurements. This was shown in \cite{Cosmography} for the case of $\alpha$, but it is generic, although currently available atomic clock tests on the stability of $\alpha$ are much more stringent than those for other constants.

The plan of the rest of this work is as follows. In Sect. \ref{method} the varying constants cosmography formalism is briefly presented. We discuss atomic clock constraints on current drift rates of dimensionless fundamental constants in Sect. \ref{clocks}, since these primarily determine the constraints on the linear terms in the cosmographic series. Constraints on the cosmographic series of $\alpha$ and $\mu$ are then reported in Sect. \ref{alphamu}, while in Sect. \ref{full} we discuss the cosmographic series of the combined $\alpha$, $\mu$, and $g_p$ data. All such constraints are fully model-independent. On the other hand, in Sect. \ref{gutmodels} we temporarily focus on cosmographic constraints under the assumption that the varying couplings arise in a class of Grand Unified Theories (GUTs) in which the variations of the three three couplings are parametrically related. Sect. \ref{forec} returns to the model-independent approach and summarizes a simple forecast exercise, which provides an estimate of the improvements expected from next-generation astrophysical facilities. Finally, Sect. \ref{concl} contains our conclusions.

\section{Varying constants cosmography formalism}\label{method}

We start by recalling the necessary mathematical formalism, trivially generalizing the one introduced for $\alpha$ in \cite{Cosmography}. One writes the relative variation of any dimensionless fundamental constant $Q$, defined in Eq. (\ref{defa}) as a Taylor series in the cosmological redshift
\be
\frac{\Delta Q}{Q}(z)=\frac{1}{Q_0}\left(\frac{dQ}{dz}\right)_0z+\frac{1}{2Q_0}\left(\frac{d^2Q}{dz^2}\right)_0z^2+\ldots\,,
\ee
where again the index $0$ denotes the present-day value. Note that by definition the constant term vanishes, so the first term in the series is the linear one. Since such a series need not converge at $z>1$ \cite{Cattoen}, one switches to the rescaled redshift $y$, cf. Eq. (\ref{defy}),
\be\label{seriesy}
\frac{\Delta Q}{Q}(y)=\frac{1}{Q_0}\left(\frac{dQ}{dy}\right)_0y+\frac{1}{2Q_0}\left(\frac{d^2Q}{dy^2}\right)_0y^2+\ldots\,;
\ee
In principle one could add cubic and higher-order terms to the series, but as briefly explored in \cite{Cosmography}, statistically, for the case of $\alpha$, the additional parameters are not needed. Therefore, in the present work, we limit ourselves to cosmographic series up to quadratic order. Moreover, we also do not consider varying constants constraints from cosmic microwave background or big bang nucleosynthesis, since such constraints are necessarily model-dependent (moreover, the former ones are far weaker than the direct constraints discussed in what follows). A brief discussion of their role can also be found in \cite{Cosmography}.

The definition of the two coefficients in the series is
\bq\label{defa1}
q_1&\equiv&\frac{1}{Q_0}\left(\frac{dQ}{dy}\right)_0=\frac{1}{Q_0}\left(\frac{dQ}{dz}\right)_0=-\frac{1}{H_0}\left(\frac{\dot Q}{Q}\right)_0\\
q_2&\equiv&\frac{1}{Q_0}\left(\frac{d^2\alpha}{dy^2}\right)_0=\frac{1}{Q_0}\left(\frac{d^2Q}{dz^2}\right)_0+ \frac{2}{Q_0}\left(\frac{dQ}{dz}\right)_0\,,\label{defa2}
\eq
where $H_0$ is the Hubble constant and the dot denotes a time derivative. For the purpose of this work, a particularly salient feature of this cosmographic expansion is that for $q_1$ the final term in brackets is the current drift rate of $Q$. For the case of $\alpha$ this can be measured in a direct and model-independent way in local laboratory experiments, which yield very stringent constraints; for $\mu$ and $g_p$, analogous indirect constraints can also be obtained, as discussed in the next section.

We can briefly illustrate the information encoded in these coefficients with two simple examples, which are representatives of the range of physically plausible varying $\alpha$ models. Firstly, consider a one-parameter redshift dependence of the form
\be
\frac{\Delta \alpha}{\alpha}(z)=\zeta_\alpha\ln{(1+z)}\,;
\ee
again in the case of $\alpha$ this is often a good low-redshift approximation for the behavior expected in string theory inspired models \cite{Damour}, in which case $\zeta_\alpha$ denotes the coupling of the dilaton to the electromagnetic sector of the theory. With these assumptions we have
\be
a_1=a_2=\zeta_\alpha\,.
\ee
In a scenario where there is experimental and observational evidence for varying $\alpha$, a consistency test for these dilaton-type class models would be the equality, within statistical uncertainties, of the linear coefficient (measured by local atomic clock experiments) and the quadratic one (measured through astrophysical observations).

Foe a second example, we take a canonical quintessence type scalar field, which is coupled to electromagnetism---as realistic quintessence fields are expected to be \cite{Carroll}---and therefore responsible both for the dynamical dark energy currently accelerating the universe and the dynamical behavior of $\alpha$. Thorough discussions of these models can be found in \cite{ROPP,Uzan}. Here we simply note the fact that in these models varying $\alpha$ and dark energy have the same underlying cause makes measurements of $\alpha$ a competitive dark energy probe. Specifically, in this case the ratio of the two cosmographic coefficients will be
\be\label{a2a1ratio}
\frac{a_2}{a_1}=1+\frac{3}{3}\Omega_mw_0+\frac{1}{2}\frac{w_0'}{1+w_0}\,,
\ee
where $\Omega_m$, $w_0$ and $w_0'$ are respectively the present-day values of the matter density (as a fraction of the critical density), of the dark energy equation of state parameter (in $c=1$ units), and of its derivative with respect to redshift. In the above we are excluding the trivial case where $w_0=-1$ exactly, since in such a case the two coefficients identically vanish.

This ratio can therefore be used as a discriminating test between thawing and freezing dark energy models, According to \cite{Caldwell}, the two classes of models can be characterized by the following two relations\footnote{Note that our present definition of $w'$ and the one used in \cite{Caldwell} differ by a minus sign.}
\bq
w_{\rm th}&=&\lambda_{\rm th}\,(1+w)\\
w_{\rm fr}&=&\lambda_{\rm fr}\,w(1+w)\,,
\eq
where both numerical coefficients are expected to be negative and of order unity
\bq
\lambda_{\rm th}&\in&[-3,-1]\\
\lambda_{\rm fr}&\in&[-3,-0.2]\,,
\eq
If so, Eq. (\ref{a2a1ratio}) can be written, in the two cases
\bq
\left(\frac{a_2}{a_1}\right)_{\rm th}&=&1+\frac{1}{2}(\lambda_{\rm th}+3\Omega_mw_0)\\
\left(\frac{a_2}{a_1}\right)_{\rm fr}&=&1+\frac{1}{2}w_0(\lambda_{\rm fr}+3\Omega_m)\,.
\eq
Assuming, for simplicity, that $\Omega_m\sim1/3$ and $w_0\sim-1$, we obtain the approximate relations
\bq
\left(\frac{a_2}{a_1}\right)_{\rm th}&\sim&\frac{1}{2}(1+\lambda_{\rm th})\\
\left(\frac{a_2}{a_1}\right)_{\rm fr}&\sim&\frac{1}{2}(1-\lambda_{\rm fr})\,.
\eq
The ratio of the cosmographic coefficients is therefore expected to be negative and positive, respectively for thawing and freezing models. The individual coefficients can have either sign, since they include a multiplicative coupling of the scalar field to the electromagnetic sector, which can itself have either sign but cancels out when the ratio is taken. The conclusion is that this ratio provides a discriminating test between the two model classes, complementing other previously discussed tests \cite{ROPP,Boas}.

\section{Atomic clock constraints}\label{clocks}

Repeated comparisons of the rates of two different atomic clocks provide a constraint on the relative shift of the corresponding characteristic frequencies. Sometimes, the same outcome can be obtained by comparing different transitions from the same atomic or molecular clock. These shifts will be proportional to a specific combination of dimensionless fundamental couplings, and thus the comparisons can be translated into a constraint of the drift of that combination. Different clock comparisons are sensitive to different products of these couplings; a combined analysis of such measurements can in principle disentangle them and lead to separate constraints on each of them.

\begin{table*}
\begin{center}
\caption{Atomic (and molecular) clock constraints of varying fundamental couplings. The third, fourth and fifth columns show the sensitivity coefficients of each frequency ratio to the various dimensionless couplings. The drift rates in the second column are given in units of $10^{-16}\,yr^{-1}$.}
\label{table1} 
\begin{tabular}{|c|c|c c c|c|}
\hline
Clocks & ${\dot \nu_{AB}}/{\nu_{AB}}$ & $\lambda_\alpha$ & $\lambda_\mu$ & $\lambda_g$ & Reference \\ 
\hline
Hg vs. Al & $0.53\pm0.79$ & -2.95 & 0.00 & 0.00 & Rosenband {\it et al.} (2008) \cite{Rosenband} \\
Dy162 vs. Dy164 & $-0.58\pm0.69$ &  1.00 & 0.00 &  0.00 & Leefer {\it et al.} (2013) \cite{Leefer} \\
Yb+(E3) vs. Yb+(E2) & $-0.012\pm0.018$ &  -6.95 & 0.00 &  0.00 & Filzinger {\it et al.} (2023) \cite{Filzinger} \\
Cs vs. Rb & $1.16\pm0.61$ &  0.49 & 0.00 & -2.00 & Abgrall {\it et al.} (2015) \cite{Abgrall} \\
Cs vs. SF${}_6$ & $-190\pm270$ &  2.83 & 0.50 & -1.27 & Shelkovnikov {\it et al.} (2008) \cite{Shelkovnikov} \\
Cs vs. H & $32\pm63$ &  2.83 & 1.00 & -1.27 & Fischer {\it et al.} (2004) \cite{Fischer} \\
Cs vs. Sr & $2.3\pm1.8$ &  2.77 & 1.00 & -1.27 & Abgrall {\it et al.} (2015) \cite{Abgrall} \\
Cs vs. Sr & $0.42\pm0.33$ &  2.77 & 1.00 & -1.27 & Schwarz {\it et al.} (2020) \cite{Schwarz} \\
Cs vs. Hg & $-3.7\pm3.9$ &  5.77 & 1.00 & -1.27 & Fortier {\it et al.} (2007) \cite{Fortier} \\
Cs vs. Yb & $0.49\pm0.36$ &  2.52 & 1.00 & -1.27 & McGrew {\it et al.} (2019) \cite{Mcgrew} \\
Cs vs. Yb+(E2) & $0.5\pm1.9$ &  1.83 & 1.00 & -1.27 & Tamm {\it et al.} (2014) \cite{Tamm} \\
Cs vs. Yb+(E3) & $-0.2\pm4.1$ &  8.83 & 1.00 & -1.27 & Huntemann {\it et al.} (2014) \cite{Huntemann} \\
Cs vs. Yb+(E3) & $0.31\pm0.34$ &  8.83 & 1.00 & -1.27 & Lange {\it et al.} (2021) \cite{Lange} \\
KRb (Molecular) & $-30\pm100$ &  0.00 & 1.00 &  0.00 & Kobayashi {\it et al.} (2019) \cite{Kobayashi} \\
\hline
\end{tabular}
\end{center}
\end{table*}
\begin{figure*}
\begin{center}
\includegraphics[width=0.33\textwidth,keepaspectratio]{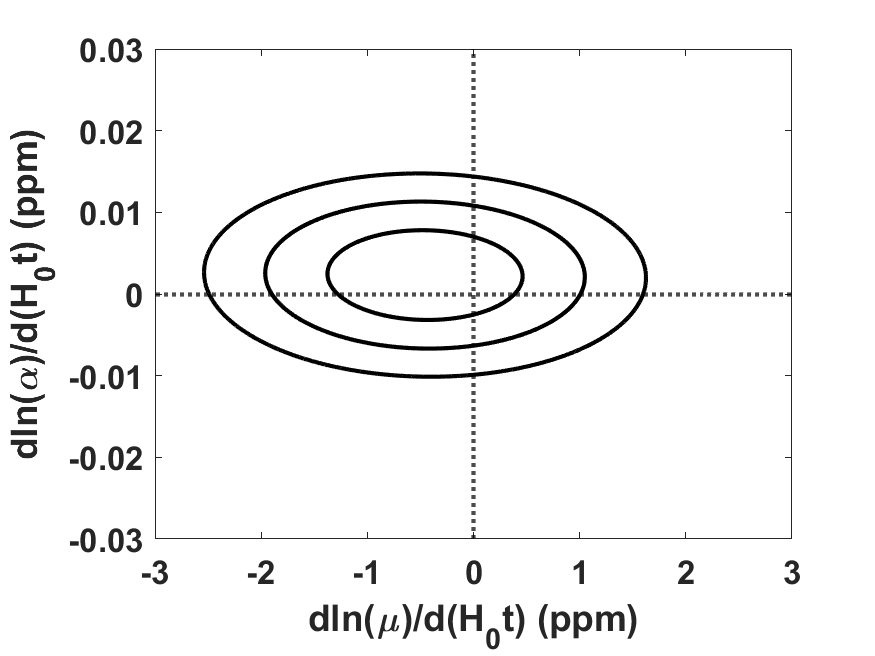}
\includegraphics[width=0.33\textwidth,keepaspectratio]{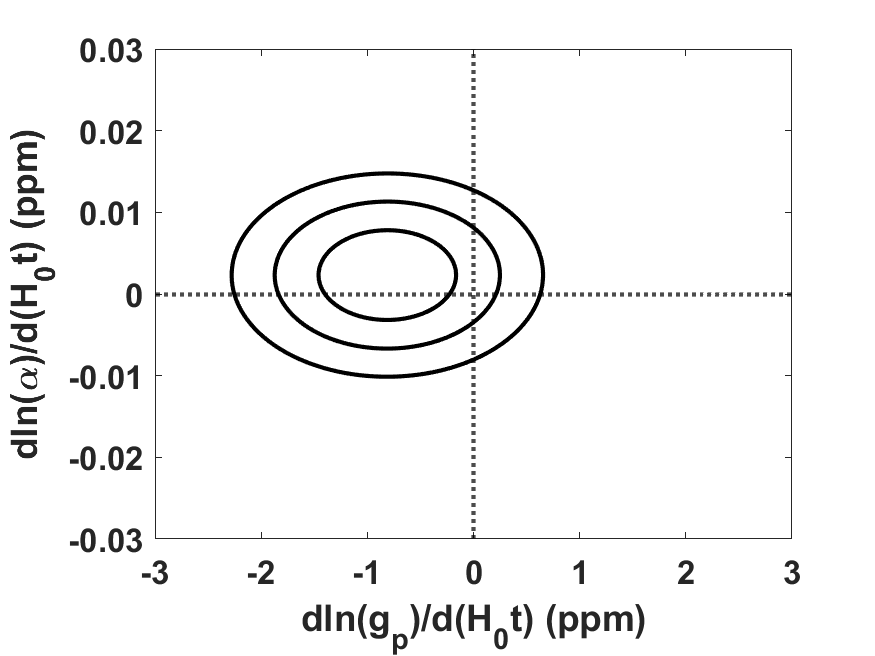}
\includegraphics[width=0.33\textwidth,keepaspectratio]{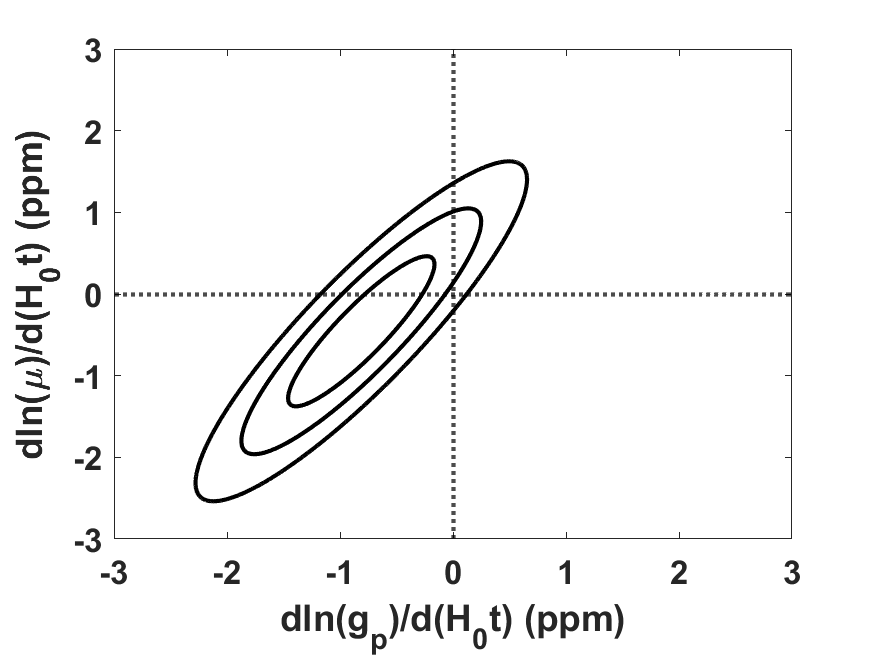}
\end{center}
\caption{\label{fig01}Atomic clock constraints on the current dimensionless drift rates of $\alpha$, $\mu$ and $g_p$,in ppm units. The one, two and three sigma confidence regions are shown in all panels, and the parameter not represented in each panel has been marginalized.}
\end{figure*}

Specifically, the ratio of two atomic clock frequencies will be proportional to
\be
\nu_{AB}=\frac{\nu_A}{\nu_B}\propto \alpha^{\lambda_\alpha} \mu^{\lambda_\mu} g_p^{\lambda_g}
\ee
where the $\lambda_i$ are the sensitivity coefficients, e.g. for $\alpha$,
\be
\lambda_\alpha=\frac{d\ln{\nu_{AB}}}{d\ln\alpha}\,,
\ee
and analogously for the other couplings. A compilation of currently available  independent measurements and their sensitivity coefficients is in Table \ref{table1} (additional details can be found in \cite{Uzan}); there are a total of 14 different measurements. The most stringent measurement is only sensitive to $\alpha$, while others are sensitive to various products of $\alpha$, $\mu$ and $g_p$.

Figure \ref{fig01} shows the results of the joint likelihood analysis of this data. Since atomic clocks constrain present-day drift rates, these constraints are usually presented in units of inverse time. Here, to facilitate comparison with astrophysical measurements we present them in dimensionless form by dividing them by the Hubble parameter, which we assume to be $H_0=70$ km/s/Mpc. (Clearly, changing this value by a few percent in either direction will not significantly impact our results.) The posterior one-sigma atomic clock constraints on each of the couplings are
\bq
\frac{d\ln{\alpha}}{H_0dt}&=&2.3\pm3.6\, ppb\\
\frac{d\ln{\mu}}{H_0dt}&=&-0.46\pm0.61\, ppm\\
\frac{d\ln{g_p}}{H_0dt}&=&-0.81\pm0.43\, ppm\,;
\eq
for convenience this and analogous subsequent values are given in units of parts per million (ppm) or, in the case of $\alpha$, parts per billion (ppb). These constraints are model-independent, other than the choice of a value of the Hubble parameter, but the latter choice is clearly subdominant with respect to other components of our study's error budget. There is no statistically significant evidence for variations. Compared to the earlier analogous study reported in \cite{Meritxell}, the constraint precision on $\alpha$ is improved by almost two orders of magnitude, while the one for $\mu$ is improved by a factor of two, and that for $g_p$ is unchanged.

\section{Cosmographic series for $\alpha$ and $\mu$}\label{alphamu}

We can now provide constraints on the individual cosmographic series for $\alpha$ and $\mu$, by combining direct astrophysical measurements of each of the two with the corresponding local constraint, discussed in the previous section.

For the fine-structure constant, $\alpha$, the astrophysical data consists of a set of 27 dedicated high-resolution spectroscopy measurements, coming from metal absorption lines in low-density absorption clouds along the line of sight of bright quasars. This includes the 21 measurements listed in Table 1 of \cite{ROPP} and a further 6 more recent ones from the Subaru telescope \cite{Cooksey}, and the HARPS \cite{Milakovic} and ESPRESSO \cite{Welsh,Murphy} spectrographs. Overall these span the redshift range $0.73\le z\le2.34$; they are expected to be more reliable than earlier measurements from archival data, and have a comparable constraining power \cite{Meritxell} to those. 

In passing, we note that we take these dedicated measurements at face value, as reported in the cited literature. In particular, when authors separately report statistical and systematic uncertainties, both of these have been included, added in quadrature. We refer the reader to \cite{Murphy} for a detailed discussion of the error budget in data reduction and analysis of these observations, specifically in the case of ESPRESSO.

\begin{figure*}
\begin{center}
\includegraphics[width=0.33\textwidth,keepaspectratio]{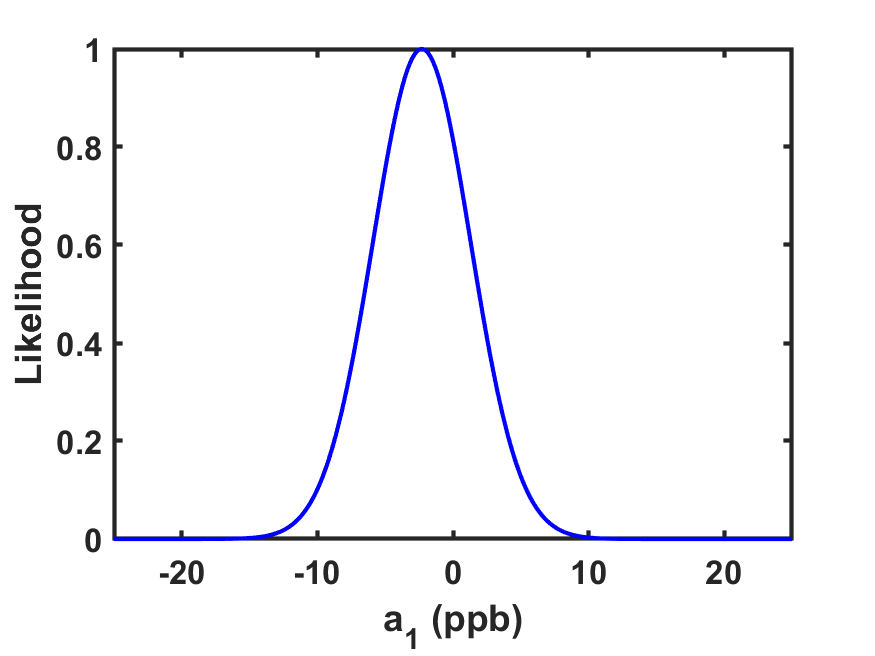}
\includegraphics[width=0.33\textwidth,keepaspectratio]{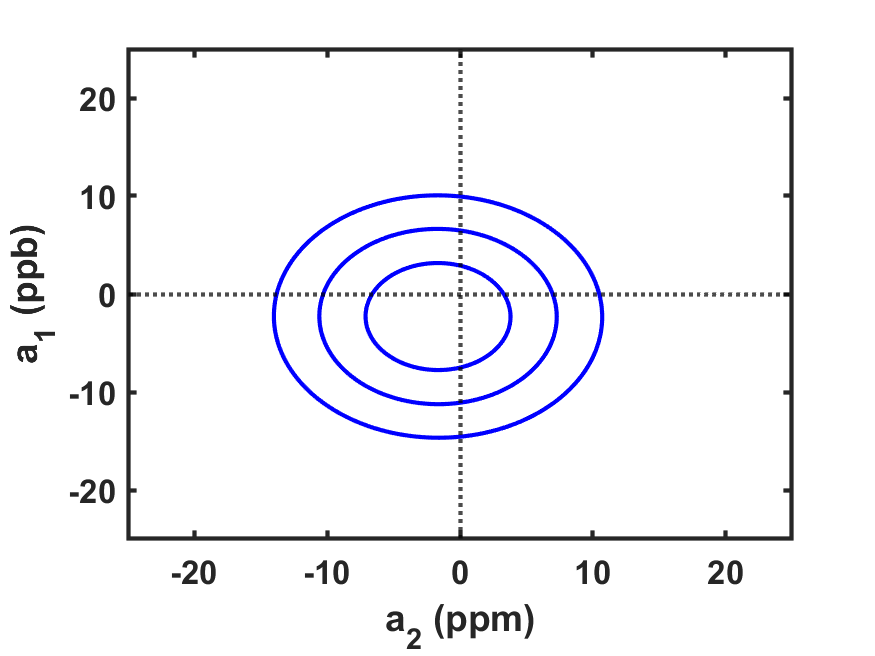}
\includegraphics[width=0.33\textwidth,keepaspectratio]{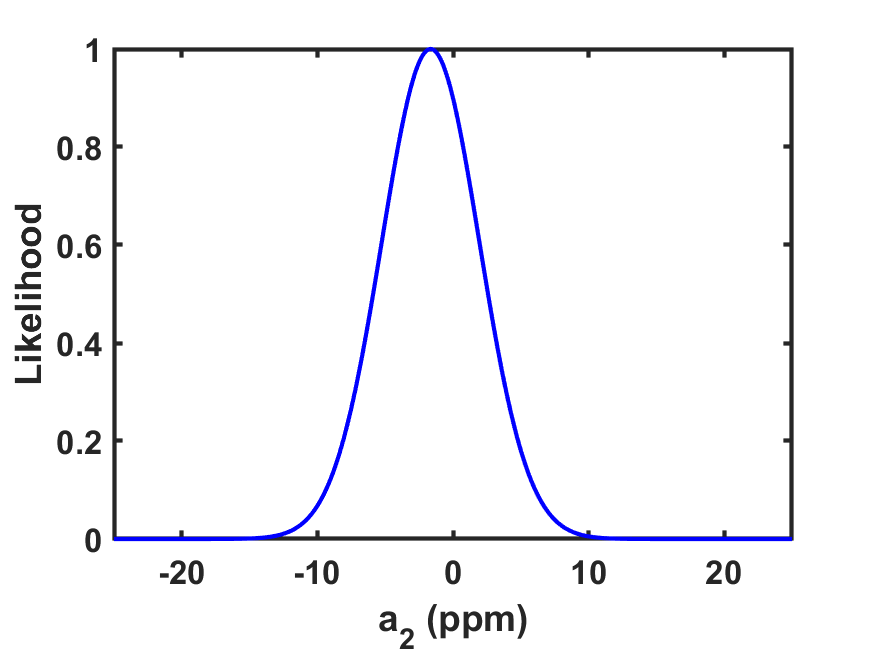}
\end{center}
\caption{\label{fig02}Constraints on the cosmographic series parameters, $a_1$ (in ppb) and $a_2$ (in ppm), for the fine-structure constant $\alpha$, from a combination of laboratory and astrophysical measurements. The middle panel shows the one, two, and three-sigma confidence levels, while the side panels show the posterior likelihoods for each parameter.}
\end{figure*}

Figure \ref{fig02} shows the results of this analysis, which leads to the constraints
\bq
a_1&=&-2.3\pm3.6\, ppb\\
a_2&=&-1.6\pm3.6\, ppm\,;
\eq
the former one is entirely driven by the atomic clocks measurement, for the aforementioned reasons. This also means that there is no significant correlation between the two parameters. Compared to the earlier analysis in \cite{Cosmography}, the constraint on $a_1$ is improved by a factor of about 4, while that of $a_2$ is not significantly changed. The constraint on the linear term of the cosmographic series is now three orders of magnitude stronger than that on the quadratic term.

For the proton-to-electron mass ratio, $\mu$, which can be measured through several vibrational and rotational molecular lines, in the optical or radio/mm bands, there is an analogous dataset of 18 measurements, of which 16 can be found in Table 2 of \cite{ROPP} and the other two are from \cite{Kanekar,Muller}. More specifically, measurements using radio/mm transitions are available at redshifts $z<1$ (there are no known high redshift targets for which the required transitions have been observed), while measurements from optical transitions are available at redshifts $z>2$ (at lower redshift the required transitions fall in the ultraviolet and can only be observed from space). Overall these span the redshift range $0.69\le z\le4.22$.

\begin{figure*}
\begin{center}
\includegraphics[width=0.33\textwidth,keepaspectratio]{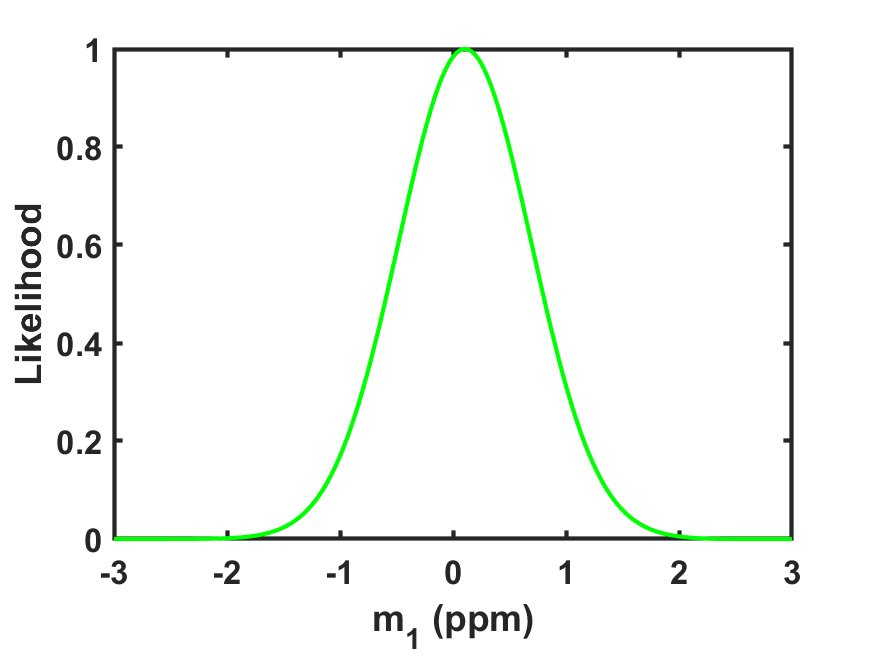}
\includegraphics[width=0.33\textwidth,keepaspectratio]{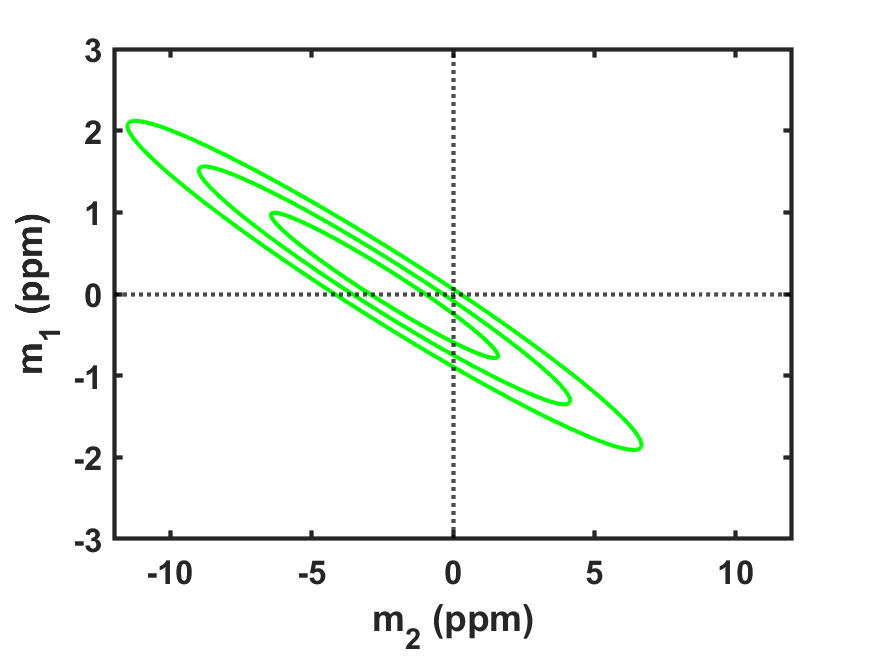}
\includegraphics[width=0.33\textwidth,keepaspectratio]{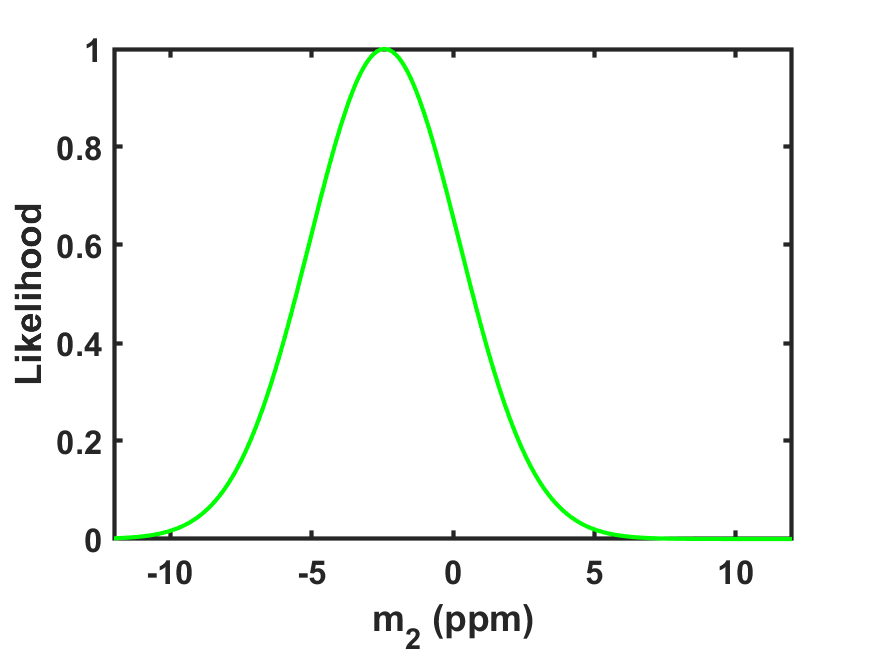}
\end{center}
\caption{\label{fig03}Constraints on the cosmographic series parameters, $m_1$ and $m_2$ (both in ppm), for the proton-to-electron mass ratio, $\mu$, from a combination of laboratory and astrophysical measurements. The middle panel shows the one, two, and three-sigma confidence levels, while the side panels show the posterior likelihoods for each parameter.}
\end{figure*}

Figure \ref{fig03} shows the results of this analysis, which leads to the constraints
\bq
m_1&=&0.1\pm0.6\, ppm\\
m_2&=&-2.4\pm2.6\, ppm\,.
\eq
The main difference between this case and the one for $\alpha$ is that here the local constraints from atomic clocks are not significantly stronger than the astrophysical ones---in fact the two are comparable. As a result, the two cosmographic parameters are constrained at about the ppm level, and they are significantly anticorrelated.

\section{General two parameter cosmographic series}\label{full}

 Analogously to the case of atomic clocks, there are several astrophysical measurements, typically done in the radio/mm part of the electromagnetic spectrum, of products of of $\alpha$, $\mu$ and $g_p$, related to the individual variations of the form
 \be\label{defqq}
\frac{\Delta Q}{Q}(z)=l_a \frac{\Delta\alpha}{\alpha}+l_m\frac{\Delta\mu}{\mu}+l_g\frac{\Delta g_p}{g_p} \,,
\ee
where the $l_i$ are the sensitivity coefficients, which differ (but are known) for each individual measurement. Here, our dataset includes a total of 35 measurements, of which 29 are listed (with the corresponding sensitivity coefficients) in Table 3 of \cite{ROPP}, while the other six are more recent and have been published in \cite{Levshakov1,Kanekar2,Gupta,Levshakov2}. Overall these span the redshift range $0.05\le z\le6.52$.

\begin{figure*}
\begin{center}
\includegraphics[width=0.33\textwidth,keepaspectratio]{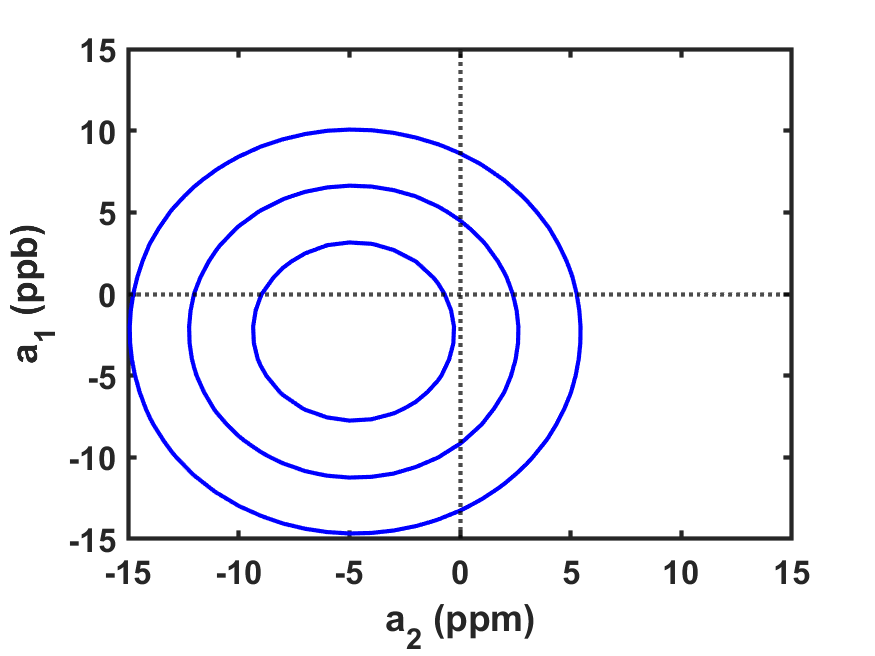}
\includegraphics[width=0.33\textwidth,keepaspectratio]{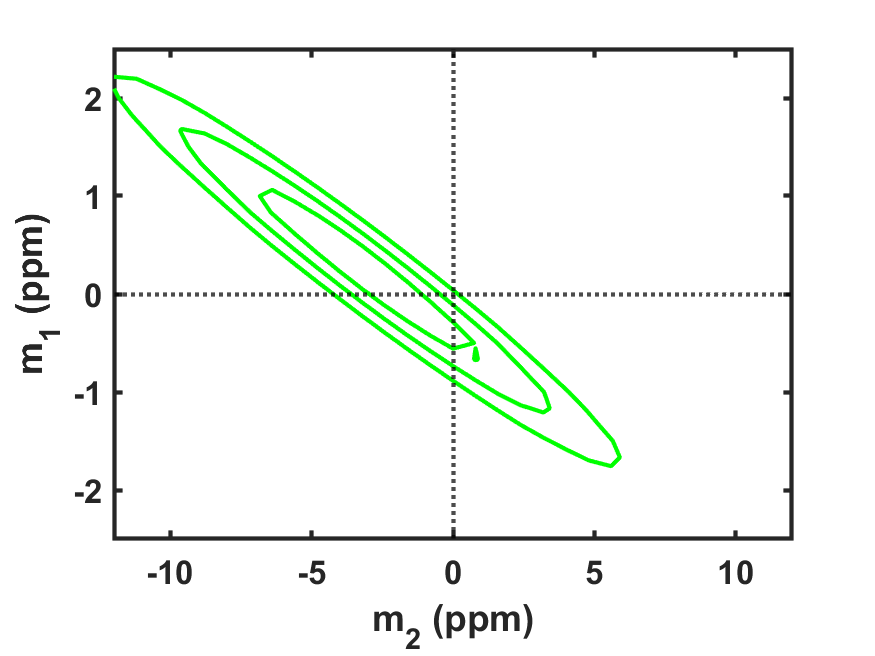}
\includegraphics[width=0.33\textwidth,keepaspectratio]{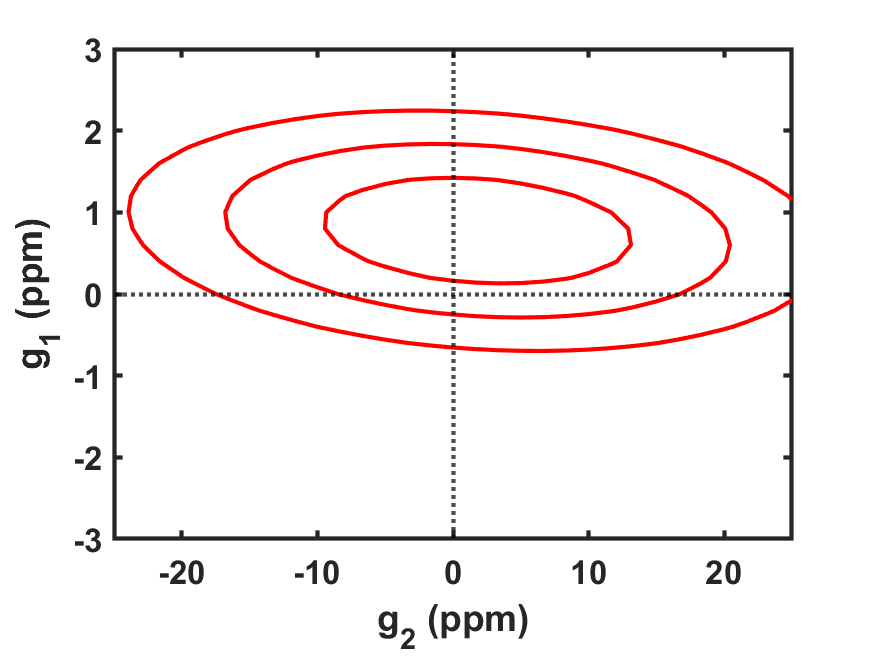}
\end{center}
\caption{\label{fig04}Constraints on the first two cosmographic series parameters, for $\alpha$, $\mu$ and $g_p$, with the four remaining parameters marginalized, from the full set of laboratory and astrophysical measurements. The one, two, and three-sigma confidence levels are shown in all cases. Notice that $a_1$ is given in ppb, while the others are in ppm.}
\end{figure*}

We can now carry out a joint likelihood analysis of all the previously described data: atomic clocks, direct measurements of $\alpha$ and $\mu$, and the combined astrophysical measurements described in the previous paragraph. In total we therefore have 94 measurements, and 6 model parameters: the first two parameters in the cosmographic series of each of the fundamental constants. The combined astrophysical measurements provide additional constraints on the cosmographic series for $\alpha$ and $\mu$, but they also lead to constraints on the cosmographic series for $g_p$.

Figure \ref{fig04} depicts the results of this analysis. Starting with $\alpha$, we obtain the one-sigma constraints
\bq
a_1&=&-2.1\pm3.8\, ppb\\
a_2&=&-5.0\pm2.8\, ppm\,;
\eq
the constraint on $a_1$ becomes marginally weaker (but is consistent with the previously reported one, being dominated by the atomic clock bound as previously discussed), while that on $a_2$ becomes tighter, and the best-fit value shifts by about one standard deviation. As for $\mu$, we now have
\bq
m_1&=&0.2\pm0.6\, ppm\\
m_2&=&-3.2\pm2.6\, ppm\,;
\eq
here the magnitude of the error bars are unchanged, with the best-fit values shifting by less than one standard deviation and in opposite directions due to the anticorrelation between the two parameters. Finally, for the proton gyromagnetic ratio we find
\bq
g_1&=&0.8\pm0.5\, ppm\\
g_2&=&1.7\pm7.5\, ppm\,;
\eq

Overall there is no statistically significant evidence for variations in any of the constants. The combined astrophysical measurements enable the constraints on $(g_1,g_2)$ but have only a mild impact on the other parameters, since their individual sensitivity is not better than a few ppm.

\section{Constraints for specific GUT models}\label{gutmodels}

The constraints presented so far are fully model-independent, in particular because they implicitly assume that any variations of $\alpha$, $\mu$ and $g_p$ are themselves independent from each other. From a phenomenological perspective this is a legitimate conservative approach. However, in specific fundamental physics paradigms this is not the case: one expects the three constants to vary in ways that are parametrically related to each other, so only one of them is truly independent. These relations will be model-dependent, but calculable in specific models. In this section We temporarily digress from the model-independent analysis in the rest of this work and consider one such example.

A convenient phenomenological paradigm for such an analysis is a broad class of GUTs \cite{Coc} which assumes that unification occurs at some not explicitly specified high energy scale, and is based on three main assumptions
\begin{itemize}
    \item the weak scale is determined by dimensional transmutation
    \item relative variations of all the Yukawa couplings are the same
    \item the variation of the couplings is driven by a scalar field \cite{Campbell}
\end{itemize}.
The consequence of these assumptions is that the relative variations of $\mu$ and $g_p$ will be related to that of $\alpha$ via
\bq
    \frac{\Delta\mu}{\mu}&=&[0.8R-0.3(1+S)]\frac{\Delta\alpha}{\alpha}\\
    \frac{\Delta g_p}{g_p}&=&[0.1R-0.04(1+S)]\frac{\Delta\alpha}{\alpha}\,,
\eq
where $R$ and $S$ are dimensionless and constant parameters, related to the quantum chromodynamics and electroweak sectors of the theory respectively. Each pair of $(R,S)$, together with the value of $\alpha$, corresponds to a specific model in this class, with $\alpha$ retaining its standard interpretation as the low-energy value measured by astrophysical observations.

This is a broad phenomenological class, and a broad range of values of $R$ and $S$, can in principle be envisaged; each pair of such values characterizes one particular model within this class. In what follows we consider three specific such models. For our purposes they are intended as representatives of the broader class, and they have been used as such in multiple works the literature. These are
\begin{itemize}
    \item The \emph{Unification} model, with $(R=36, S=160)$, has been argued to include the most typical parameter values in this model class \cite{Coc};
    \item The \emph{Dilaton} model, with $(R=109.4, S=0)$, which draws inspiration from string-theory type scalar fields \cite{Nakashima};
    \item The \emph{UV cutoff} model, with $(R=-183, S=22.5)$, in which one assumes that the eponymous cutoff is cosmologically varying \cite{Lee}.
\end{itemize}

\begin{figure*}
\begin{center}
\includegraphics[width=0.33\textwidth,keepaspectratio]{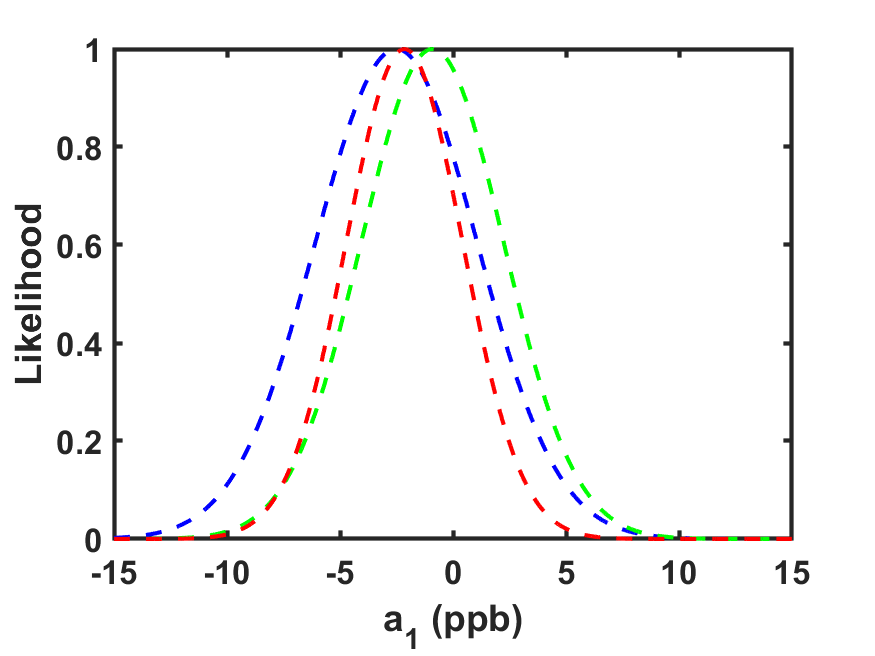}
\includegraphics[width=0.33\textwidth,keepaspectratio]{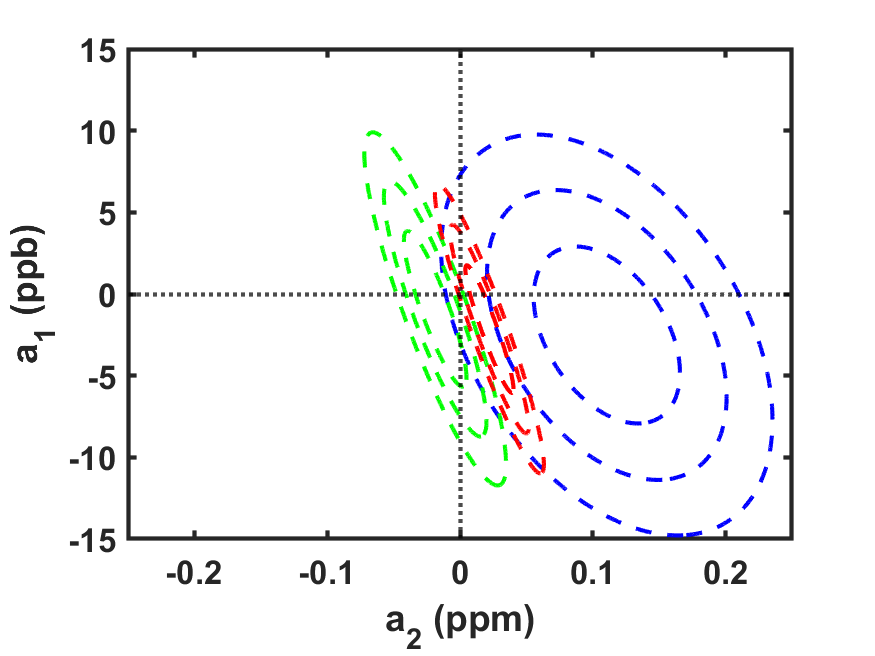}
\includegraphics[width=0.33\textwidth,keepaspectratio]{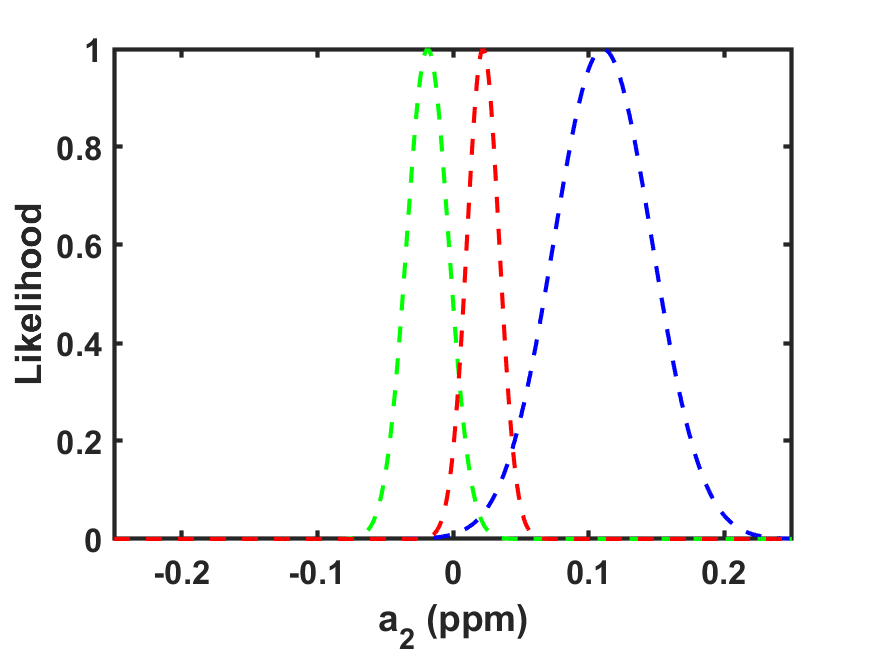}
\end{center}
\caption{\label{fig05}Constraints on the first two cosmographic series parameters for $\alpha$, $a_1$ (in ppb) and $a_2$ (in ppm), from all the available atomic clock and astrophysical data, assuming the the Unification, Dilaton and UV cutoff models, shown in blue, green and red respectively. The middle panel shows the one, two, and three-sigma confidence levels, while the side panels show the posterior likelihoods for each parameter.}
\end{figure*}

By relating the behavior of the three couplings these models reduce our parameter space to the $\alpha$ cosmographic series. This can be envisaged as choosing theoretical priors which relate the various types of measurements. On the other hand, the model dependence of the relation implies that the derived constraints will be different in each model. This is illustrated in Figure \ref{fig05}, and the constraints are also summarized in the first three rows od Table \ref{table2}.

The comparison confirms a model dependence of the results, although it is noteworthy that this predominantly applies to the second cosmographic coefficient, $a_2$; for the first coefficient this dependence is noticeably milder, for the aforementioned reason: $a_1$ is strongly constrained by the local atomic clock results. Nevertheless, an anticorrelation between the two parameters is now visible for all three models, showing that the $\mu$ and $g_p$ data has a significant impact in the overall analysis. It is also worthy of note that the constraints on $a_2$ are improved by approximately two orders of magnitude, the reason being that the numerical factor relating the relative variations of $\mu$ and $g_p$ with those of $\alpha$ is of this order, for our choices of $(R,S)$. The unification model is the one which yields the weakest constraints; its best-fit value value for $a_2$ differs from the null result by about 2.7 standard deviations, which again is not statistically significant.

\begin{table}
\begin{center}
\caption{One-sigma constraints on the first two cosmographic series parameters for $\alpha$, $a_1$ (in ppb) and $a_2$ (in ppm). for the three specific GUT models discussed in the text.}
\label{table2} 
\begin{tabular}{|c c c c c |}
\hline
Model & $R$ & $S$ & $a_1$ (ppb) & $a_2$ (ppm) \\ 
\hline
Unification & 36 & 160 & $-2.5\pm3.6$ & $0.11\pm0.04$ \\
Dilaton & 109.4 & 0 & $-0.9\pm3.2$ & $-0.02\pm0.02$ \\
UV cutoff & -183 &  22.5 & $-2.1\pm2.6$ & $0.02\pm0.01$ \\
\hline
\end{tabular}
\end{center}
\end{table}

\section{Outlook: A simple next-generation forecast}\label{forec}

We now briefly report on a forecasting exercise, estimating how the current constraints may be improved by relevant next-generation facilities, under construction or planned. Since our focus is on astrophysical measurements, we will conservatively assume that the sensitivity of current atomic clock measurements is unchanged. On the other hand, for the astrophysical measurements we assume that all systems considered in the previous sections are re-observed at higher sensitivity (by a factor to be specified below), with the further assumption of a fiducial model where there are no variations.

This assumption of a uniform gain in sensitivity is of course simplistic, but at least it only relies on targets which already have provided measurements and can be re-measured at higher sensitivity. It does not account for the possibility of measurements in new targets---either to be discovered, or already known but too faint to provide a good measurement in a reasonable integration time with current facilities. Moreover, the assumption of a fiducial model without variations is also a conservative one, both in the obvious theoretical sense and because it is the case for which degeneracies between the cosmographic parameters are maximal.

For the optical measurements, we assume a gain in sensitivity of a factor of 5, which is trivially expected from ESPRESSO to ANDES \cite{HIRES,ANDES} by simply considering the increased collecting area of the ELT and assuming that the two spectrographs have similar performance. For the radio/mm measurements, we envisage an ALMA2040-like facility, currently under discussions as part of ESO's  Expanding Horizons initiative\footnote{\url{https://next.eso.org/}} \cite{ALMA}, and specifically considering a gain in sensitivity of a factor of 5 or 10.

With the above caveats, the results are shown in Table \ref{table3}, both for the two facilities separately (meaning that only one of the astrophysical data sets is assumed to have an improved sensitivity) and for the case of a joint analysis (where both have sensitivity improvements). There is no significant gain on $a_1$, due to the extremely tight atomic clock measurement already available, For $\alpha$ the gains are larger in the optical, while the opposite happens for $\mu$ and $g_p$. It's also manifest that there is an important synergy between the
two facilities.

It's also worthy of note that in the case where both facilities have improved sensitivities by a factor of 5 one never gains a factor of 5 on the constraints on the cosmographic parameters, but the reasons for this are twofold and clear. The first reason is that the forecast assumes no change in the lab experiments (which is also conservative, as they will improve), and the second one is that there are limiting degeneracies between these parameters, Nevertheless, for $a_2$ and $g_2$ the gains are very large. Unsurprisingly, we also see that there is a significant difference between the factor of 5 or 10 sensitivity gain for an ALMA2040-like facility.

\begin{table*}
\begin{center}
\caption{One sigma constraints on the various cosmographic coefficients, from the current data discussed in Sect. \ref{full}, for ANDES-like and ALMA2040-like datasets (as described in the main text), and the combination of the latter two. Values of $a_1$ are given in ppb, all other cosmographic coefficients are given in ppm. The gain in sensitivity for each coefficient, with respect to the current constraints, is also reported.}
\label{table3} 
\begin{tabular}{|c|c c|c c|c c|c c|c c|c c|}
\hline
Dataset & $a_1$ & Gain & $a_2$ & Gain & $m_1$ & Gain & $m_2$ & Gain & $g_1$ & Gain & $g_2$ & Gain  \\ 
\hline
Current & 3.8 & - & 2.8 & - & 0.58 & - & 2.6 & - & 0.47 & - & 7.5 & - \\
\hline
ANDES (5x) & 3.8 & 1.00 & 0.71 & 3.94 & 0.50 & 1.16 & 2.2 & 1.18 & 0.41 & 1.15 & 4.5 & 1.67 \\
\hline
ALMA2040 (5x) & 3.7 & 1.03 & 1.1 & 2.55 & 0.33 & 1.76 & 2.0 & 1.30 & 0.41 & 1.15 & 2.8 & 2.68 \\
Joint & 3.7 & 1.03 & 0.60 & 4.67 & 0.25 & 2.32 & 1.3 & 2.00 & 0.40 & 1.18 & 2.0 & 3.75 \\
\hline
ALMA2040 (10x) & 3.6 & 1.06 & 0.56 & 5.00 & 0.21 & 2.76 & 0.86 & 3.02 & 0.37 & 1.27 & 1.7 & 4.41 \\
Joint & 3.6 & 1.06 & 0.44 & 6.36 & 0.19 & 3.05 & 0.82 & 3.17 & 0.34 & 1.38 & 1.4 & 5.36 \\
\hline
\end{tabular}
\end{center}
\end{table*}

\section{Conclusions}\label{concl}

We have built upon the recent work in \cite{Cosmography} to carry out an updated and broader cosmographic analysis of the stability of three dimensionless fundamental couplings: the fine-structure constant, proton-to-electron mass ratio, and proton gyromagnetic ratio. The three are relevant because they are comparatively easy to measure, on their own (for the first two) or in various multiplicative combinations, both in local experiments using atomic clocks and in astrophysical observations.

A noteworthy feature of the cosmographic approach to varying couplings is that atomic clock measurements directly constrain the first cosmographic coefficient, although astrophysical measurements are also sensitive to it. In the case of the fine-structure constant, for which particularly sensitive atomic clock measurements are available, these lead to an extremely stringent ppb level constraint on $a_1$. For the other couplings, atomic clock constraints are comparatively weaker, and their constraining power on the first cosmographic coefficients ($m_1$ and $g_1$) is comparable to that of astrophysical measurements, at slightly under the ppm level.

On the other hand, the second cosmographic coefficients are only constrained by astrophysical measurements, and current data also restricts them to ppm level. In this work we have not considered higher-order terms in the series; according to the analysis in \cite{Cosmography} these terms are not statistically warranted by the $\alpha$ data considered therein, and we have verified that this result also applies to the extended data sets considered in the present work. Intuitively, this may be anticipated bearing in mind that any putative coupling variations must be cosmologically slow and, for most physically realistic models, limited to the matter era, so a series truncated at quadratic order should suffice.

One other difference with respect to the earlier work is that in the present one we have not considered constraints coming from the cosmic microwave background or big bang nucleosynthesis, since both of these are model-dependent. Additionally, for the case of the cosmic microwave background, such constraints are so weak as to have no impact on the results we presented. On the other hand, a more detailed analysis of the impact of nucleosynthesis constraints on varying couplings is ongoing and will be reported elsewhere.

Overall, we find no statistically significant for variations, whether we rely on the fully model-independent approach or we assume that the variations of the three couplings are parametrically related as predicted in a particular class of GUT models. We emphasize that constraints at the ppm level (or better) are extremely stringent when compared with other contemporary probes of the dark universe. The closest comparable approach is dynamical dark energy, for which the analogous dimensionless parameter (viz. the dark energy equation of state) can only be constrained to the level of a few percent.

Moreover, in many (though not all) theoretical paradigms varying fundamental couplings and dynamical dark energy are actually related, so observational constraints on the former lead to constraints on the latter \cite{ROPP,Uzan}. Section \ref{method} provides a simple illustration of this point: the ratio of the first two cosmographic coefficients in the $\alpha$ series provides a discriminating test between models, and in particular can distinguish between canonical freezing and thawing dark energy models. More broadly, ratios of cosmographic coefficients of different couplings can also distinguish between models, e.g. they may ultimately lead to tight constraints on the values of $(R,S)$ in the class of GUT models discussed in Sect. \ref{gutmodels}.
 
Finally, we note that much progress is expected in the sensitivity of the relevant datasets. Local laboratory experiments relying on Th-229 nuclear transitions have the potential to improve the sensitivity to $\alpha$ by at least two orders of magnitude \cite{Thorium1,Thorium2,Thorium3,Beeks}, which would improve constraints on $a_1$ by a similar amount. As for astrophysical observations, new facilities such as the ELT can significantly improve existing measurements \cite{HIRES,ANDES}, provided one builds upon recently learned lessons \cite{Schmidt1,Schmidt2}, and emerging new facilities such as ALMA2040 can also have a major impact in the field. Ultimately, as previously discussed in \cite{Peebles}, a detailed survey of the behavior of fundamental dimensionless couplings, addressing both possible time (redshift) and environmental dependencies, may prove to be a more fruitful way to explore the dark side of the universe than traditional approaches.

\section*{Acknowledgements}

This work was financed by Portuguese funds through FCT (Funda\c c\~ao para a Ci\^encia e a Tecnologia) in the framework of the project 2022.04048.PTDC (Phi in the Sky, DOI 10.54499/2022.04048.PTDC). CJM also acknowledges FCT and POCH/FSE (EC) support through Investigador FCT Contract 2021.01214.CEECIND/CP1658/CT0001 (DOI 10.54499/2021.01214.CEECIND/CP1658/CT0001).

Discussions on this topic with Mar Artigas, Celia Clerfeuille, Noelia Vadillo and other members of the Phi in the Sky team are gratefully acknowledged.

\bibliographystyle{model1-num-names}
\bibliography{cosmography}
\end{document}